\documentclass[aps,twocolumn,epsfig,prb]{revtex4}

\newcommand{\bsigma}{\mbox{\boldmath $\sigma$}}

\usepackage{amsmath}
\usepackage{graphicx}

\def\nn{\nonumber}

\begin{document}

\title{Curvature induced optical phonon frequency shift 
in metallic carbon nanotubes} 

\author{K.~Sasaki$^{a}$, R.~Saito$^{a}$, G.~Dresselhaus$^{b}$,
M.~S.~Dresselhaus$^{c,d}$, 
H.~Farhat$^{e}$, and J.~Kong$^{d}$
}

\affiliation{$^{a}$Department of Physics, Tohoku University and CREST, JST, 
Sendai, 980-8578, Japan}
\affiliation{$^{b}$Francis Bitter Magnet Laboratory, 
$^{c}$Department of Physics, 
$^{d}$Department of Electrical Engineering and Computer Science, 
$^{e}$Department of Materials Science and Engineering,
Massachusetts Institute of Technology, Cambridge, MA 02139-4307}


\date{\today}
 
\begin{abstract}
 The quantum corrections to 
 the frequencies of the $\Gamma$ point
 longitudinal optical (LO) and transverse optical (TO) phonon modes 
 in carbon nanotubes 
 are investigated theoretically.
 The frequency shift and broadening of the TO phonon mode 
 strongly depend on the curvature effect 
 due to a special electron-phonon coupling in carbon nanotubes, 
 which is shown by the Fermi energy dependence of the frequency shift 
 for different nanotube chiralities.
 It is also shown that 
 the TO mode near the $\Gamma$ point decouples 
 from electrons due to local gauge symmetry and 
 that a phonon mixing between LO and TO modes 
 is absent due to time-reversal symmetry.
 Some comparison between theory and experiment is presented.
\end{abstract}

\maketitle

\section{introduction}

In the Raman spectra of a single wall carbon
nanotube (SWNT), the two in-plane optical phonon
modes, that is, the longitudinal optic (LO) and
transverse optic (TO) phonon modes at the $\Gamma$
point in the two-dimensional Brillouin zone (2D
BZ), which are degenerate in graphite and
graphene, split into two peaks, $G^+$ and $G^-$
peaks, respectively.~\cite{s617, d1018, f1020}
The splitting of the two
peaks for SWNTs is inversely proportional to the
square of the diameter $d_t$ of SWNTs due to the
curvature effect, in which $G^+$ does not change
with changing $d_t$, but the $G^-$ frequency decreases
with decreasing $d_t$.~\cite{m871,saito07}
In particular, for
metallic SWNTs, the $G^-$ peaks appear at a lower
frequency than the $G^-$ peaks for semiconducting
SWNTs with a similar diameter.~\cite{w699} 
The spectra of $G^-$
for metallic SWNTs show a much larger spectral
width than that for semiconducting SWNTs. 
Further,
the spectral $G^-$ feature shows an asymmetric lineshape as a
function of frequency which is known as
the Breit-Wigner-Fano (BWF) lineshape.~\cite{a807} 
The origin of the BWF lineshape is considered to be
due to the interaction of discrete phonon states with
continuous free electron states.

%

It has been widely accepted 
that the frequency shift of the $G$-band 
is produced by the electron-phonon (el-ph)
interaction.~\cite{piscanec04,lazzeri06prl,ishikawa06,popov06} 
An optical phonon 
changes into an electron-hole pair
as an intermediate state 
by the el-ph interaction. 
This process gives the phonon a self-energy.
The phonon self-energy is 
sensitive to the Fermi energy, $E_{\rm F}$.
In the case of graphite intercalation compounds 
in which the charge transfer of an electron 
from a dopant to the graphite layer 
can be controlled by the doping atom and its concentration, 
Eklund {\it et al.} observed 
a shift of the $G$-band frequency 
with an increase of the spectral width.~\cite{N14} 
In this case the frequency shifted spectra show that 
not only the LO mode but also the TO mode
are shifted in the same fashion by a dopant.
For a graphene monolayer,
Lazzeri {\it et al.} calculated 
the $E_{\rm F}$ dependence of the shift 
of the G-band frequency.~\cite{lazzeri06prl}
The LO mode softening in metallic SWNTs was shown by
Dubay {\it et al.},~\cite{dubay02,dubay03} 
on the basis of density functional theory.
Recently Nguyen {\it et al.}~\cite{nguyen07} 
and Farhat {\it et al.}~\cite{l1208}
observed the phonon softening effect of SWNTs 
as a function of $E_{\rm F}$ by field
effect doping and electro-chemical doping,
respectively, 
and their results clearly show that the LO
phonon modes become soft as a function of $E_{\rm F}$.
Ando discussed the phonon softening for metallic SWNTs
as a function of the $E_{\rm F}$ position, 
in which the phonon softening occurs for the LO phonon mode 
and for a special range of $E_{\rm F}$, 
that is, for 
$|E_{\rm F}|<\hbar\omega_{\rm LO}/2$.~\cite{ando08}


In this paper, 
we show that the $\Gamma$ point TO phonon mode becomes hard
when $|E_{\rm F}| \alt \hbar\omega_{\rm TO}/2$
and has a considerable broadening 
for metallic zigzag nanotubes.
The occurrence of the phonon hardening for the TO mode
is due to the curvature effect,
a special character of the electron-phonon coupling,
and a basic consequence of second-order perturbation theory.
We show using a gauge symmetry argument that
the electrons completely decouple from
the TO mode near the $\Gamma$ point.
Besides, we show that for a chiral nanotube, 
both the LO and TO modes are softened 
due to the fact that the direction of the TO phonon vibration
is not parallel to the nanotube circumferential
direction.~\cite{reich01}
Another interest
of ours is the mixing of LO and TO phonons to form
degenerate phonon frequencies. When the LO phonon
mode becomes soft, a crossing of the LO mode with the TO
mode occurs at a certain value of $E_{\rm F}$. 
For such a mode crossing, 
we should generally consider the electron-phonon coupling
for degenerate phonon modes to sense the
crossing or anti-crossing of the two phonon
frequencies as a function of $E_{\rm F}$. 
We will show by an analytical calculation
that 
there is no mixing between the LO and TO phonon modes 
for any case due to time reversal symmetry.

The organization of the paper is as follows. 
In Sec.~\ref{sec:ps} 
we show our method of calculation and 
present the results for armchair and metallic zigzag SWNTs. 
In Sec.~\ref{sec:chi}
using effective mass theory, 
we show how the el-ph interaction 
depends on the chiral angles of SWNTs, 
and in Sec.~\ref{sec:dis}
a discussion based on gauge symmetry and
time reversal symmetry for the el-ph coupling
is given.
In Sec.~\ref{sec:sum},
a comparison with the experiments and summary are given.

\section{phonon frequency shift}\label{sec:ps}

\begin{figure}[htbp]
 \begin{center}
  \includegraphics[scale=0.45]{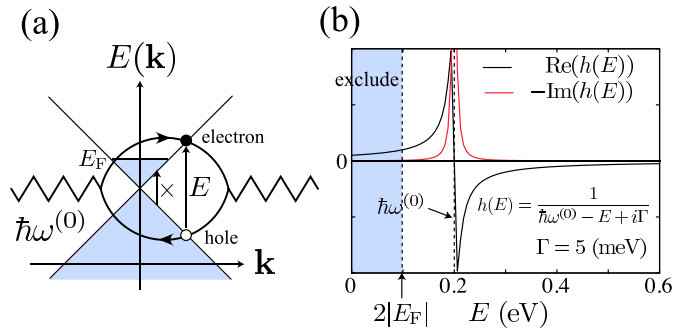}
 \end{center}
 \caption{(color online)
 (a) 
 An intermediate electron-hole pair state that 
 contributes to the energy shift of the optical phonon modes 
 is depicted.
 A phonon mode is denoted by a zigzag line and an electron-hole
 pair is represented by a loop.
 The low energy electron-hole pair satisfying 
 $0 \le E \le 2|E_{\rm F}|$ is forbidden  at zero temperature
 by the Pauli principle.
 (b) The energy correction to the phonon energy 
 by an intermediate electron-hole pair state, 
 especially the sign of the correction,
 depends on the energy of the intermediate state as $h(E)$.
 }
 \label{fig:process}
\end{figure}

The frequency shift of the $\Gamma$-point LO and TO phonon modes 
for metallic SWNTs
is calculated by second-order perturbation theory.
The phonon energy including the el-ph interaction
becomes $\hbar \omega_\lambda=
\hbar \omega_\lambda^{(0)}+\hbar \omega_\lambda^{(2)}$
($\lambda={\rm LO},{\rm TO}$)
where $\omega_\lambda^{(0)}$ 
is the original phonon frequency
without the el-ph interaction
and $\hbar \omega_\lambda^{(2)}$ is given by
\begin{align}
 \hbar \omega_\lambda^{(2)} = 
 & 2 \sum_{\bf k}
 \frac{|\langle {\rm eh}({\bf k})|{\cal H}_{\rm int}|\omega_\lambda
 \rangle|^2}{\hbar\omega_\lambda^{(0)}-(E_e({\bf k})-E_h({\bf
 k}))+i\Gamma_\lambda}  \nn \\
 & \times 
 \left(f(E_h({\bf k})-E_{\rm F})-f(E_e({\bf k})-E_{\rm F}) \right).
 \label{eq:omega_2}
\end{align}
The factor 2 in Eq.~(\ref{eq:omega_2}) comes from spin degeneracy.
$\hbar \omega_\lambda^{(2)}$ is the quantum correction 
to the phonon energy due to the
electron-hole pair creation as shown in Fig.~\ref{fig:process}(a).
In Eq.~(\ref{eq:omega_2}),
$\langle {\rm eh}({\bf k})|{\cal H}_{\rm int}|\omega_\lambda \rangle$
is the matrix element for creating an electron-hole pair
at momentum ${\bf k}$
by the el-ph interaction, ${\cal H}_{\rm int}$,
$E_e({\bf k})$ ($E_h({\bf k})$) 
is the electron (hole) energy
and $\Gamma_\lambda$ is the decay width.
In Fig.~\ref{fig:process}(a), 
an intermediate electron-hole pair state
that has the energy of $E=E_e({\bf k})-E_h({\bf k})$
is shown.
We need to sum ($\sum_{\bf k}$) 
over all possible intermediate electron-hole pair states
in Eq.~(\ref{eq:omega_2}), 
which can have a much larger energy than the phonon
($E \gg \hbar \omega_\lambda^{(0)}$).

Since
$\langle {\rm eh}({\bf k})|{\cal H}_{\rm int}|\omega_\lambda \rangle$
is a smooth function of
$E=E_e({\bf k})-E_h({\bf k})$
in the denominator of Eq.~(\ref{eq:omega_2}),
the contribution to $\hbar \omega_\lambda^{(2)}$
from an electron-hole pair depends on its energy.
In Fig.~\ref{fig:process}(b), 
we plot the real part and imaginary part of 
the denominator of Eq.~(\ref{eq:omega_2}),
$h(E)=1/(\hbar \omega^{(0)}-E+i\Gamma)$ as a function of $E$
in the case of $\hbar \omega^{(0)}=0.2$ eV and $\Gamma=5$ meV.
Here ${\rm Re}(h(E))$ is a positive (negative) value
when $E < \hbar \omega^{(0)}$ ($E > \hbar \omega^{(0)}$) and
the lower (higher) energy electron-hole pair
makes a positive (negative) contribution to 
$\hbar \omega_\lambda^{(2)}$.
Moreover,
an electron-hole pair satisfying $E<2|E_{\rm F}|$ 
can not contribute to the energy shift
(shaded region in Fig.~\ref{fig:process}(a) and (b)) 
because of the Fermi distribution function $f(E)$ in
Eq.~(\ref{eq:omega_2}).
Thus, the quantum correction to the phonon energy 
by an intermediate electron-hole pair 
can be controlled by changing
the Fermi energy, $E_{\rm F}$.
For example, when $|E_{\rm F}| = \hbar \omega^{(0)}/2$,
then $\hbar \omega_\lambda^{(2)}$ takes a minimum value 
at zero temperature
since all positive contributions to $\hbar \omega_\lambda^{(2)}$
are suppressed in Eq.~(\ref{eq:omega_2}). 
Since ${\rm Re}(h(E)) \approx -1/E$ for $E \gg \hbar \omega^{(0)}$,
all high energy intermediate states 
contribute to phonon softening 
if we include all the electronic states in the system.
Here we introduce a cut-off energy at $E_c=0.5$ eV 
as $\sum_{\bf k}^{E_e({\bf k})< E_c}$
in order to avoid such a large energy shift in Eq.~(\ref{eq:omega_2}).
The energy shift due to the high-energy intermediate states
($\sum_{\bf k}^{E_e({\bf k})> E_c}$)
can be neglected by renormalizing $\hbar \omega^{(0)}$ 
so as to reproduce the experimental results 
of Raman spectra~\cite{l1208}
since the contribution from $E_e({\bf k})> E_c$ 
just gives a constant energy shift to $\hbar \omega^{(2)}$.
We have checked that the present results do not depend on 
the selection of the cut-off energy since
$E_c$ is much larger than $\hbar \omega^{(0)}$.

${\rm Im}(h(E))$ is nonzero only very close to 
$E=\hbar \omega^{(0)}$,
which shows that the phonon can resonantly 
decay into an electron-hole pair 
with the same energy.
It is noted that when $|E_{\rm F}| > \hbar \omega^{(0)}/2$,
$\Gamma_\lambda \approx 0$ at zero temperature 
while $\Gamma_\lambda$ 
may take a finite value at a finite temperature.
In this paper, we calculate 
$\Gamma_\lambda$ self-consistently
by calculating
$\Gamma_\lambda= - {\rm Im} (\hbar \omega_\lambda^{(2)})$
in Eq.~(\ref{eq:omega_2}).

\begin{figure}[htbp]
 \begin{center}
  \includegraphics[scale=0.5]{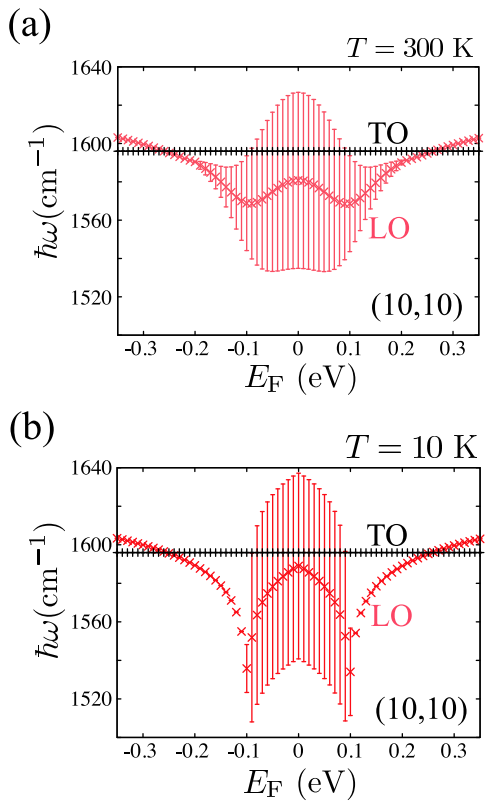}
 \end{center}
 \caption{(color online)
 The $E_{\rm F}$ dependence of the LO (red curve) and TO (black
 curve) phonon energy in the case of the $(10,10)$ armchair nanotube. 
 (a) is taken at room temperature and (b) is at 10 K.
 Only the energy of the LO mode is shifted,
 with the TO mode frequency being independent of $E_{\rm F}$.
 The decay width ($\Gamma_\lambda$) is plotted as an error-bar.
 }
 \label{fig:arm}
\end{figure}

In Fig.~\ref{fig:arm},
we show calculated results for
$\hbar \omega_\lambda$ as a function of $E_{\rm F}$
for a $(10,10)$ armchair nanotube.
Here we take 1595 ${\rm cm}^{-1}$ and 1610 ${\rm cm}^{-1}$
for $\hbar \omega_\lambda^{(0)}$ of
the TO and LO modes, respectively.
The energy bars denote $\Gamma_\lambda$ values.
We have used the extended tight-binding scheme 
to calculate $E_e({\bf k})$, $E_h({\bf k})$, and
the electron wavefunction for
$\langle {\rm eh}({\bf k})|{\cal H}_{\rm int}|\omega_\lambda
\rangle$.~\cite{samsonidze04}
As for the el-ph matrix element,~\cite{jiang05prb}
we adopted the deformation potential derived
on the basis of density-functional
theory by Porezag {\it et al}.~\cite{porezag95}
We show the resulting $\hbar \omega_\lambda$ 
as a function of $E_{\rm F}$
at the room temperature ($T=300$ K) and $T=10$ K
in Fig.~\ref{fig:arm}(a) and (b), respectively.
It is shown that the TO mode does not exhibit any energy change 
while the LO mode shows an energy shift and broadening.
As we have mentioned above, the minimum energy is realized at 
$|E_{\rm F}| = \hbar \omega^{(0)}/2$ ($\approx 0.1$ eV).
There is a local maximum for the spectral peak 
at $|E_{\rm F}| = 0$.
The broadening for the LO mode occurs 
within $|E_{\rm F}| \alt \hbar \omega^{(0)}/2$
for the lower temperature 
while the broadening has a tail at room temperature 
for $|E_{\rm F}| \agt \hbar \omega^{(0)}/2$.

\begin{figure}[htbp]
 \begin{center}
  \includegraphics[scale=0.45]{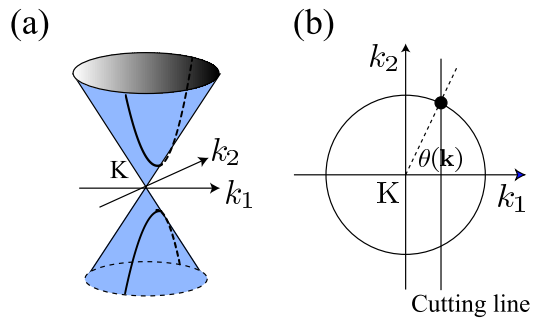}
 \end{center}
 \caption{
 (a) Cutting line near the K-point.
 The $k_1$ ($k_2$) axis is selected 
 as the nanotube circumferential (axis) direction. 
 The amplitude for an electron-hole pair creation 
 depends strongly on the relative position of the cutting line
 from the K-point.
 (b) If the cutting line crosses the K-point, then
 the angle $\theta({\bf k})$ ($\equiv \arctan(k_2/k_1)$)
 takes $\pi/2$ ($-\pi/2$) values for $k_2 > 0$ ($k_2 < 0$).
 In this case, the LO mode strongly couples to an electron-hole pair,
 while the TO mode is decoupled from the electron-hole pair
 according to Eq.~(\ref{eq:coupling}).
 }
 \label{fig:scattering}
\end{figure}

A continuous model for electrons in a carbon nanotube is adopted 
in this paper to explain the lack of an energy shift of the TO modes 
for armchair nanotubes.
As we will show in Sec.~\ref{sec:chi},
the el-ph matrix element for the electron-hole pair creation 
by the LO and TO modes is given by
\begin{align}
 \begin{split}
  &\langle {\rm eh}({\bf k})|{\cal H}_{\rm int}|\omega_{\rm LO} \rangle
  =-ig u \sin \theta({\bf k}), \\
  &\langle {\rm eh}({\bf k})|{\cal H}_{\rm int}|\omega_{\rm TO} \rangle
  =-ig u \cos \theta({\bf k}).
 \end{split}
 \label{eq:coupling}
\end{align}
where $u$ is the phonon amplitude, 
and $g$ is the el-ph coupling constant.
Here $\theta({\bf k})$ is the angle 
for the polar coordinate around the K (or K') point
in the 2D BZ, in which
a ${\bf k}=(k_1,k_2)$ point on a cutting line 
for a metallic energy
subband is taken.
The $k_1$ ($k_2$) axis is taken in the direction of 
the nanotube circumferential (axis) direction
(see Fig.~\ref{fig:scattering}). 
Equation~(\ref{eq:coupling}) shows that
$\langle {\rm eh}({\bf k})|{\cal H}_{\rm int}|\omega_\lambda \rangle$
depends only on $\theta({\bf k})$ but not on $|{\bf k}|$,
which means that the dependence 
of this matrix element
on $E$ is negligible.
Since the armchair nanotube is free from 
the curvature effect,~\cite{saito92prb}
the cutting line for its metallic energy band
lies on the $k_2$ axis.
Thus, we have $\theta({\bf k})=\pi/2$ ($-\pi/2$)
for the metallic energy subband 
which has $k_1=0$ and $k_2>0$ ($k_2<0$).
Then, Eq.~(\ref{eq:coupling}) tells us that
only the LO mode couples to an electron-hole pair
and the TO mode is not coupled
to an electron-hole pair
for armchair SWNTs.

\begin{figure}[htbp]
 \begin{center}
  \includegraphics[scale=0.6]{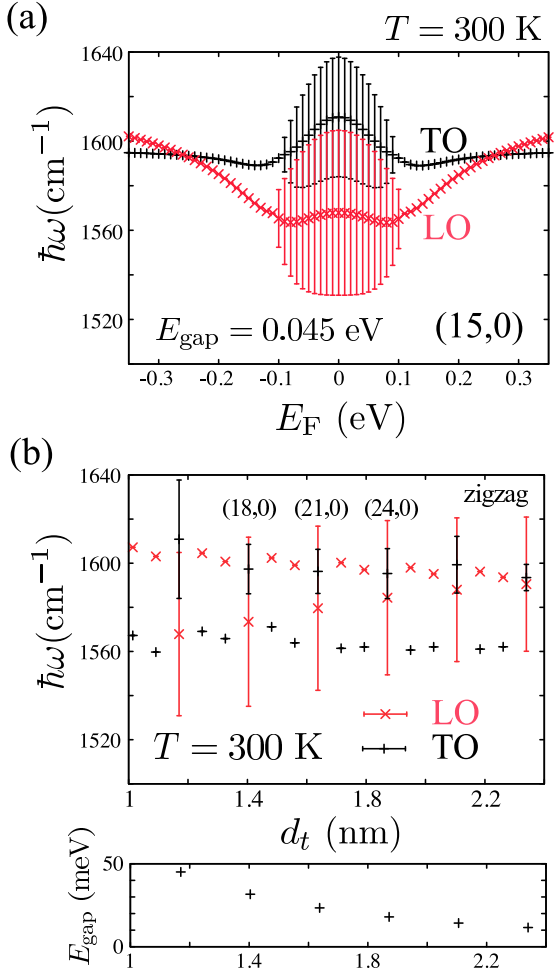}
 \end{center}
 \caption{(color online)
 (a) The $E_{\rm F}$ dependence of the LO (red curve) 
 and TO (black curve) phonon frequency for a $(15,0)$ zigzag nanotube. 
 Not only the frequency of the LO mode but also that of the TO mode 
 is shifted due to the curvature effect.
 (b) The diameter $d_t$ dependence of the phonon frequency
 for zigzag nanotubes, including zigzag semiconducting tubes.
 $E_{\rm gap}$ denotes the curvature-induced mini energy gap.
 }
 \label{fig:zig}
\end{figure}

In Fig.~\ref{fig:zig}(a),
we show calculated results for
$\hbar \omega_\lambda$ as a function of $E_{\rm F}$
for a $(15,0)$ metallic zigzag nanotube.
In the case of zigzag nanotubes,
not only the LO mode but also the TO mode
couples with electron-hole pairs.
The spectrum peak position for the TO mode
becomes harder for $E_{\rm F}=0$,
since ${\rm Re}(h(E))$ for 
$E < \hbar \omega_{\rm TO}$
contributes to a positive frequency shift.
It has been shown theoretically~\cite{saito92prb} and
experimentally~\cite{ouyang01} that even for
``metallic'' zigzag nanotubes
a finite curvature opens a small energy gap.
When the curvature effect is taken into account,
the cutting line does not lie on the K-point, 
but is shifted from the $k_2$ axis.
In this case,
$\cos \theta({\bf k})=k_1/(k_1^2 + k_2^2)^{1/2}$
is nonzero for the lower energy intermediate 
electron-hole pair states due to $k_1 \ne 0$.
Thus, the TO mode can couple to 
the low energy electron-hole pair which
makes a positive energy contribution to the phonon energy shift.
The high energy electron-hole pair 
still decouples to the TO mode
since $\cos\theta({\bf k}) \to 0$ for $|k_2|\gg |k_1|$.
Therefore, when $|E_{\rm F}| \alt \hbar \omega^{(0)}_{\rm TO}/2$,
then $\hbar \omega_{\rm TO}$ increases by a larger amount than 
$\hbar \omega_{\rm LO}$.
The TO mode for the small diameter zigzag nanotubes 
couples strongly with an electron-hole pair 
because of the stronger curvature effect.
In Fig.~\ref{fig:zig}(b),
we show the diameter ($d_t$) dependence of the $\hbar \omega_\lambda$
of zigzag nanotubes for $E_{\rm F}=0$
not only for metallic SWNTs but also for semiconducting SWNTs.
In the case of the semiconducting nanotubes,
the LO (TO) mode appears around 1600 (1560) ${\rm cm}^{-1}$
without any broadening.
Only the metallic zigzag nanotubes show an energy shift,
and the energy of the LO (TO) mode decreases (increases)
as compared to the semiconducting tubes.
In the lower part of Fig.~\ref{fig:zig}(b),
we show a curvature-induced energy gap $E_{\rm gap}$
as a function of $d_t$.
The results show that 
higher (lower) energy electron-hole pairs contribute effectively 
to the LO (TO) mode softening (hardening) in metallic nanotubes.
In the case of semiconducting nanotubes,
we may expect that there is a softening 
for the LO and TO modes according to Eq.~(\ref{eq:coupling}).
However, the softening is small as compared with that of the metallic
nanotubes because the energy of intermediate electron-hole pair states
is much larger than $\hbar \omega_\lambda^{(0)}$
in this case.

\section{Chirality dependence of the electron-phonon
 interaction}\label{sec:chi}

\begin{figure}[htbp]
 \begin{center}
  \includegraphics[scale=0.45]{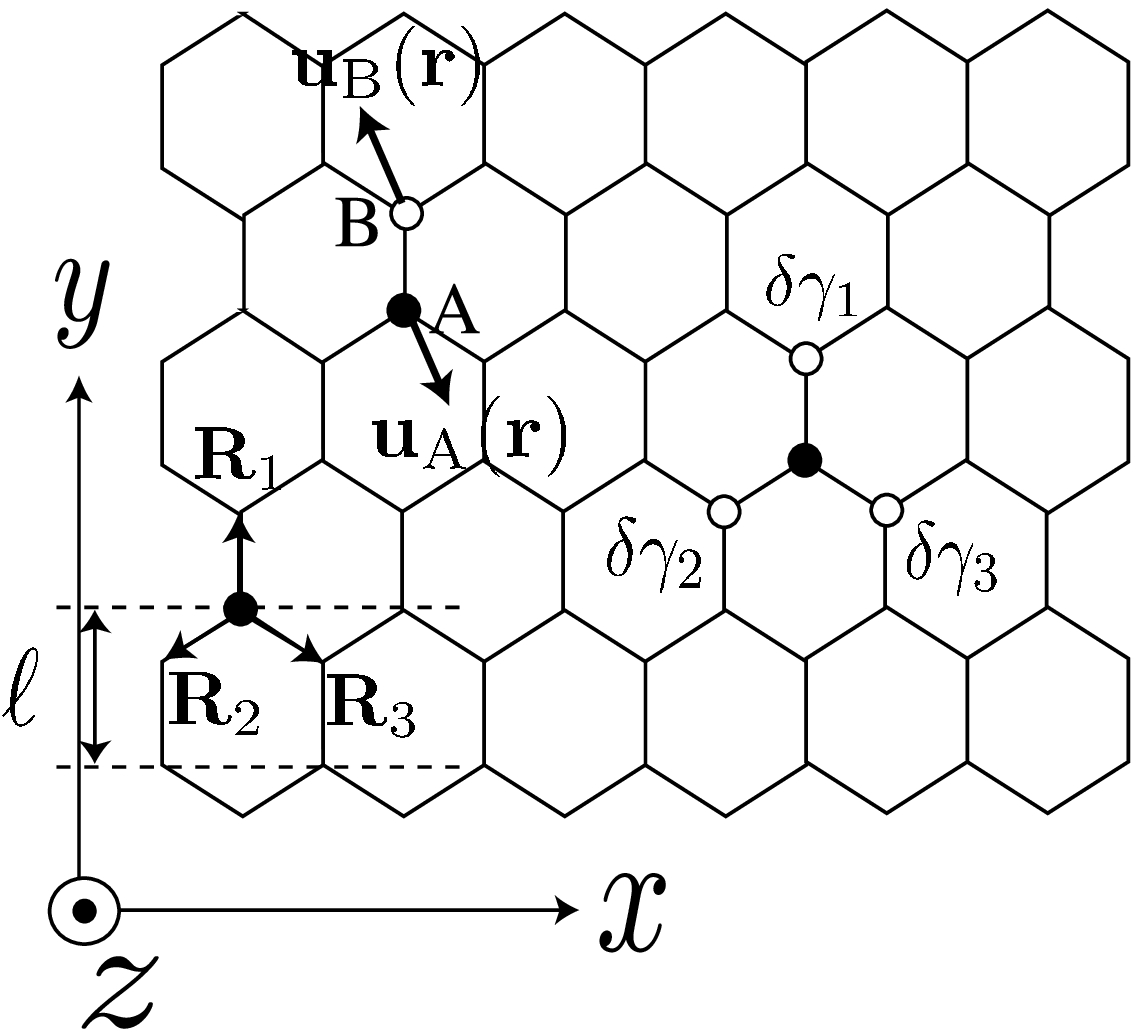}
 \end{center}
 \caption{A hexagonal unit cell of graphene consists of 
 {\rm A} (closed circle) and {\rm B} (open circle) sublattices.
 ${\bf R}_a$ ($a=1,2,3$)
 are vectors pointing to the nearest-neighbor  
 {\rm B} sites from an {\rm A} site (${\bf R}_1=a_{\rm cc}{\bf e}_y$,
 ${\bf R}_2=-(\sqrt{3}/2)a_{\rm cc}{\bf e}_x -(1/2)a_{\rm cc}{\bf e}_y$,
 and 
 ${\bf R}_3=(\sqrt{3}/2)a_{\rm cc}{\bf e}_x -(1/2)a_{\rm cc}{\bf e}_y$).
 Local modulations of the hopping integral are defined by 
 $\delta \gamma^a_0(\mathbf{r})$ ($a=1,2,3$).
 The modulation is given by optical phonon modes as
 $\delta \gamma^a_0 = (g/\ell) {\bf u}({\bf r}) \cdot {\bf R}_a$
 where ${\bf u}({\bf r})$ 
 ($={\bf u}_{\rm A}({\bf r})-{\bf u}_{\rm B}({\bf r})$) 
 is a relative displacement vector of a {\rm B} site
 relative to the nearest {\rm A} site.
 }
 \label{fig:graphene}
\end{figure}

Here, we derive Eq.~(\ref{eq:coupling})
on the basis of a continuous model for $\pi$-electrons 
near the K-point in graphene.
In a continuous model,
the local modulation of the hopping integral
due to lattice vibrations appears 
as a deformation-induced gauge field, 
${\bf A}({\bf r})=(A_x({\bf r}),A_y({\bf r}))$,
in the Weyl equation.~\cite{sasaki05}
The Weyl equation for $\pi$-electrons with energy eigenvalue $E$ 
is written by
\begin{align}
 v_F \bsigma \cdot ({\hat{\bf p}}+{\bf A}({\bf r}))
 \Psi({\bf r})=E\Psi({\bf r}),
 \label{eq:weyl}
\end{align}
where
$v_F$ ($\equiv \gamma_0\ell/\hbar$) is the Fermi velocity
and $\gamma_0$ ($\approx 2.7$eV) 
is the nearest-neighbor hopping integral,
$\ell\equiv (3/2)a_{\rm cc}$, $\hat{\bf p}=-i\hbar\nabla$
is the momentum operator, and
$\bsigma=(\sigma_x,\sigma_y)$ is the Pauli matrix.
${\bf A}({\bf r})$ is given in Eq.~(3) of Ref.~\onlinecite{sasaki06jpsj}
by a small change $\delta \gamma_a({\bf r})$ ($a=1,2,3$) 
of the hopping integral from $-\gamma_0$
(see Fig.~\ref{fig:graphene}),
as
\begin{align}
 \begin{split}
  & v_F A_x({\bf r}) = \delta \gamma^1_0({\bf r})
  - \frac{1}{2} \left( \delta \gamma^2_0({\bf r}) +
  \delta \gamma^3_0({\bf r}) \right), \\
  & v_F A_y({\bf r}) = \frac{\sqrt{3}}{2} 
  \left( \delta \gamma^2_0({\bf r}) -
  \delta \gamma^3_0({\bf r}) \right).
 \end{split}
 \label{eq:A}
\end{align}
Here
$\delta \gamma^a_0({\bf r})$ for the LO and TO modes is given by
$\delta \gamma^a_0({\bf r})=(g/\ell) {\bf u}({\bf r}) \cdot {\bf R}_a$
where ${\bf R}_a$ denotes the nearest-neighbor vectors 
(Fig.~\ref{fig:graphene})
and ${\bf u}({\bf r})$ is the relative displacement vector 
of a {\rm B} site from an {\rm A} site 
(${\bf u}({\bf r})={\bf u}_{\rm B}({\bf r})-{\bf u}_{\rm A}({\bf r})$)
and $g$ is the el-ph coupling constant.
We rewrite Eq.~(\ref{eq:A}) as
\begin{align}
 v_{\rm F} (A_x({\bf r}),A_y({\bf r})) = g (u_y({\bf r}),-u_x({\bf r})),
 \label{eq:optigauge}
\end{align}
where $u_i({\bf r}) \equiv {\bf u}({\bf r}) \cdot {\bf e}_i$, ($i=x,y$),
and ${\bf R}_1-({\bf R}_2+{\bf R}_3)/2=\ell {\bf e}_y$ 
and $\sqrt{3}/2({\bf R}_2-{\bf R}_3)=-\ell {\bf e}_x$
have been used (see Fig.~\ref{fig:graphene}).
Then, 
the el-ph interaction 
for an in-plane lattice distortion ${\bf u}({\bf r})$
can be rewritten as the
vector product of $\bsigma$ and
${\bf u}({\bf r})$,~\cite{ishikawa06}
\begin{align}
 {\cal H}_{\rm int} = 
 v_{\rm F} \bsigma \cdot {\bf A}({\bf r})
 =g(\bsigma\times{\bf u}({\bf r}))\cdot {\bf e}_z.
 \label{eq:H_int}
\end{align}

We consider the LO and TO phonon modes with ${\bf q}={\bf 0}$
(i.e., $\Gamma$-point).
Then, an electron-hole pair is excited by 
a constant ${\bf u}=(u_x,u_y)$.
The el-ph matrix element for the electron-hole pair generation 
is given by
\begin{align}
 &\langle {\rm eh}({\bf k})|{\cal H}_{\rm int}|\omega \rangle
 = \int 
 \Psi^*_{c,{\bf k}}({\bf r}) {\cal H}_{\rm int}
 \Psi_{v,{\bf k}}({\bf r}) d^2{\bf r}
 \nonumber \\
  = &\frac{g}{2}
 \begin{pmatrix}
  e^{+i\frac{\Theta({\bf k})}{2}} &
  e^{-i\frac{\Theta({\bf k})}{2}}
 \end{pmatrix}
 \begin{pmatrix}
  0 & u_y + i u_x \cr
  u_y - i u_x & 0 
 \end{pmatrix} 
 \begin{pmatrix}
  e^{-i\frac{\Theta({\bf k})}{2}} \cr 
  -e^{+i\frac{\Theta({\bf k})}{2}}
 \end{pmatrix},
 \label{eq:mat_elph}
\end{align}
where $\Psi_{c,{\bf k}}({\bf r})$ 
($\Psi_{v,{\bf k}}({\bf r})$)
denotes an energy eigenstate of 
$v_F \bsigma \cdot {\bf p}$ in the conduction (valence)
energy band with energy eigenvalue $E=v_{\rm F}|{\bf p}|$
($E=-v_{\rm F}|{\bf p}|$),
\begin{align}
 \begin{split}
  &
  \Psi_{c,{\bf k}}({\bf r})
  = \frac{e^{i{\bf k}\cdot {\bf r}}}{\sqrt{2S}}
  \begin{pmatrix}
   e^{-i\Theta({\bf k})/2} \cr
   e^{+i\Theta({\bf k})/2}
  \end{pmatrix},
  \\
  &
  \Psi_{v,{\bf k}}({\bf r})
  = \frac{e^{i{\bf k}\cdot {\bf r}}}{\sqrt{2S}}
  \begin{pmatrix}
   e^{-i\Theta({\bf k})/2} \cr
   -e^{+i\Theta({\bf k})/2}
  \end{pmatrix},
 \end{split}
\end{align}
where $S$ denotes the total area of the system, 
$k_x - i k_y\equiv |{\bf k}| e^{-i\Theta({\bf k})}$.

We first consider the case of 
a zigzag nanotube in Fig.~\ref{fig:graphene}.
Then, we denote $x$ ($y$) 
as a coordinate around (along) the axis
(so that $\Theta({\bf k})=\theta({\bf k})$),
and $u_x({\bf r})$ ($u_y({\bf r})$)
are assigned to the TO (LO) phonon mode.
The corresponding ${\bf A}({\bf r})$ 
for $u_x({\bf r})$ and $u_y({\bf r})$
is $A_y({\bf r})$ and $A_x({\bf r})$, respectively.
By calculating Eq.~(\ref{eq:mat_elph}) for the TO mode with 
$(u_x,u_y) = (u,0)$  and for the LO mode with
$(u_x,u_y) = (0,u)$, we get Eq.~(\ref{eq:coupling}).
Next, we consider the case of an armchair nanotube.
Then, $x$ ($y$) is the coordinate along (around) the axis 
(so that $\Theta({\bf k})=\theta({\bf k})+\pi/2$), and 
$u_x({\bf r})$ ($u_y({\bf r})$)
is assigned to the LO (TO) phonon mode.
The direction of the gauge field ${\bf A}({\bf r})$
is perpendicular to the phonon eigenvector ${\bf u}({\bf r})$
and the LO mode shifts the wavevector around the tube axis,
which explains how
the LO mode may induce a dynamical energy band-gap 
in metallic nanotubes.~\cite{dubay02} 
By calculating Eq.~(\ref{eq:mat_elph}) for the TO mode with 
$(u_x,u_y) = (0,u)$  and for the LO mode with
$(u_x,u_y) = (u,0)$, we get Eq.~(\ref{eq:coupling}), too.
Equation~(\ref{eq:coupling}) is valid 
regardless of the tube chirality if the 
phonon eigenvector of the LO (TO) phonon mode is 
in the direction along (around) the tube axis.
This is because ${\hat {\bf p}}$ and ${\bf u}({\bf r})$ 
are transformed in the same way
as we change the chiral angle.
As a result, 
there would be no chiral angle dependence for the el-ph coupling
in Eq.~(\ref{eq:coupling}).

\begin{figure}[htbp]
 \begin{center}
  \includegraphics[scale=0.5]{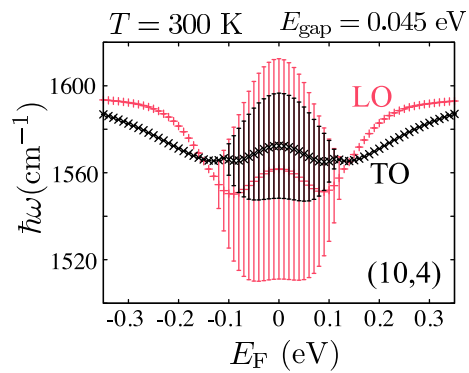}
 \end{center}
 \caption{(color online)
 The $E_{\rm F}$ dependence of the LO (red curve) and TO (black
 curve) phonon energy for the $(10,4)$ metallic chiral nanotube. 
 $E_{\rm gap}$ denotes the curvature-induced mini energy gap which has
 the same value as that for the $(15,0)$ tube in Fig.~\ref{fig:zig}(a).
 The difference between the behavior of 
 the $(10,4)$ and $(15,0)$ nanotubes
 comes from the fact that 
 the phonon eigenvector of the LO (TO) mode is not pointing 
 along the tube
 axis in the case of the $(10,4)$ tube. 
 }
 \label{fig:chi}
\end{figure}

However, 
the phonon eigenvector depends on the chiral angle.
Reich {\it et al.} reported that, for a chiral nanotube, 
atoms vibrate along the direction of the carbon-carbon bonds
and not along the axis or the circumference.~\cite{reich01}
In the case of a chiral nanotube,
the phonon eigenvector may be written as
\begin{align}
 \begin{pmatrix}
  u_{\rm TO} \cr u_{\rm LO}
 \end{pmatrix}
 =
 \begin{pmatrix}
  \cos \phi & \sin \phi \cr
  - \sin \phi & \cos \phi
 \end{pmatrix}
 \begin{pmatrix}
  u_1 \cr u_2
 \end{pmatrix},
 \label{eq:chi}
\end{align}
where $u_1$ ($u_2$) is in the direction 
around (along) a chiral tube axis,
and $\phi$ is the angle difference
between the axis and the vibration.
This modifies Eq.~(\ref{eq:coupling}) as
\begin{align}
 \begin{split}
  &\langle {\rm eh}({\bf k})|{\cal H}_{\rm int}|\omega_{\rm LO} \rangle
  =-ig u \sin (\theta({\bf k})+\phi), \\
  &\langle {\rm eh}({\bf k})|{\cal H}_{\rm int}|\omega_{\rm TO} \rangle
  =-ig u \cos (\theta({\bf k})+\phi).
 \end{split}
 \label{eq:coupling_chi}
\end{align}
In Fig.~\ref{fig:chi},
we show numerical results for
$\hbar \omega_\lambda$ as a function of $E_{\rm F}$
for a $(10,4)$ metallic chiral nanotube
for $T=300$ K.
The energy difference between the minimum point at 
$E_{\rm F} = \hbar \omega^{(0)}/2$ and at $E_{\rm F} = 0$
shows that the LO mode couples more strongly to the low energy 
electron-hole pair than the TO mode.
Since the curvature-induced mini energy gap for $(10,4)$
and $(15,0)$ tubes are almost the same ($E_{\rm gap}=0.045$ eV), 
we may expect a similar energy shift of the LO and TO modes.
However, the results of the $(15,0)$ zigzag tube
show that the TO mode couples more strongly for the low energy 
electron-hole pair than the LO mode.
This is explained not by Eq.~(\ref{eq:coupling}) but
by Eq.~(\ref{eq:coupling_chi}) with appropriate $\phi$
($\phi\approx 23.4^{\circ}$), which
is given by the phonon eigenvector calculation.
Since the chiral angle for $(10,4)$ is $16.1^{\circ}$,
$\phi$ is not directly related to the chiral angle.
The identification of $\phi$ in Eq.~(\ref{eq:coupling_chi})
as a function of chirality would be useful to compare theoretical
results and experiments, 
which will be explored in the future.

\section{Discussion}\label{sec:dis}

The phonon frequencies with some broadening 
as shown in Fig.~\ref{fig:arm} are not 
directly related to the Raman spectra
of the G-band 
but relate to the phonon density of states 
of the LO and TO phonons at $q=0$. 
In previous papers,~\cite{r824}
we have shown that the G-band intensity 
depends on chiral angle in which for zigzag nanotubes
the LO (TO) phonon mode is strong (suppressed) 
while for armchair nanotubes
the TO (LO) phonon mode is strong (comparable). 
This chirality dependence of the G-band intensity 
is also observed in single nanotube Raman spectroscopy.~\cite{yu01,b1016} 
This chirality dependence of the G-band mode intensity
is exactly opposite to the chirality dependence of the phonon-softening in
which the TO phonon mode is suppressed in armchair nanotubes.
These observations 
clearly show that the corresponding el-ph interaction for
phonon-softening and Raman processes are independent of each other.

\begin{figure}[htbp]
 \begin{center}
  \includegraphics[scale=0.5]{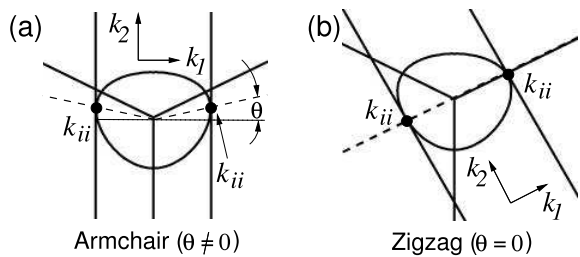}
 \end{center}
 \caption{Trigonal warping effect of an equi-energy surface in the 
2D Brillouin zone and cutting lines for (a) armchair and 
(b) zigzag nanotubes. In the case of (a) armchair nanotubes,
the $\theta$ value defined from the $k_1$ axis in Fig.~\ref{fig:graphene} 
is not zero while in the case of (b) zigzag nanotubes, 
$\theta$ is always zero, 
which explains the chirality dependence of 
the relative Raman intensity of LO and TO phonon modes.   
 }
 \label{fig:tri}
\end{figure}

In the Raman process, the phonon is emitted by scattering an
electron (or an exciton) from one conduction band state to another
conduction band state, while in the phonon softening, the electron is
scattered (or excited) from a valence band state to a conduction band state.
Thus the corresponding matrix element for the Raman process is
expressed by substituting $\Psi_{c,{\bf k}}({\bf r})$ in 
Eq.~(\ref{eq:mat_etoe}) for $\Psi_{v,{\bf k}}({\bf r})$ in 
Eq.~(\ref{eq:mat_elph}) as follows, 
\begin{align}
 &\langle {\rm e}({\bf k}),\omega|{\cal H}_{\rm int}|{\rm e}({\bf k}) \rangle
 = \int 
 \Psi^*_{c,{\bf k}}({\bf r}) {\cal H}_{\rm int}
 \Psi_{c,{\bf k}}({\bf r}) d^2{\bf r}
 \nonumber \\
  = &\frac{g}{2}
 \begin{pmatrix}
  e^{+i\frac{\Theta({\bf k})}{2}} &
  e^{-i\frac{\Theta({\bf k})}{2}}
 \end{pmatrix}
 \begin{pmatrix}
  0 & u_y + i u_x \cr
  u_y - i u_x & 0 
 \end{pmatrix} 
 \begin{pmatrix}
  e^{-i\frac{\Theta({\bf k})}{2}} \cr 
  e^{+i\frac{\Theta({\bf k})}{2}}
 \end{pmatrix}.
 \label{eq:mat_etoe}
\end{align}
Then the electron-phonon matrix element 
for the Raman scattering process
becomes   
\begin{align}
 \begin{split}
  & \langle {\rm e}({\bf k}), \omega_{\rm LO}|
  {\cal H}_{\rm int}|{\rm e}({\bf k}) \rangle
  =g u \cos \theta({\bf k}), \\
  & \langle {\rm e}({\bf k}), \omega_{\rm TO}|
  {\cal H}_{\rm int}|{\rm e}({\bf k}) \rangle
  =-g u \sin \theta({\bf k}).
 \end{split}
 \label{eq:coupling2}
\end{align}
It is stressed again that
the el-ph interactions of Eqs.~(\ref{eq:coupling})
and ~(\ref{eq:coupling2}) are for 
phonon-softening and Raman intensity, respectively
and that they are independent of each other.
When we consider the Raman scattering processes for carbon nanotubes,  
the ${\bf k}$ vector which is relevant to the resonance Raman
process is the ${\bf k}$ vector at the van Hove singular point, 
${\bf k}_{ii}$, which corresponds to the touching points 
of the equi-energy surface to the cutting line 
(one dimensional Brillouin zone plotted in the 2D BZ)~\cite{s617,t1008} 
which are shown in Fig.~\ref{fig:tri}(a) and (b) 
for zigzag and armchair nanotubes, 
respectively. 

In the case of the zigzag nanotubes, 
the ${\bf k}_{ii}$ point corresponds to $\theta = 0$ 
which is not the chiral angle, 
but is defined in Fig.~\ref{fig:scattering}(b). 
Thus from Eq.~(\ref{eq:coupling2}), 
we see that the TO mode is not excited. 
On the other hand, 
in the case of the armchair nanotube, 
the value of $\theta$ is not zero 
because of the trigonally warped equi-energy surface as shown in
Fig.~\ref{fig:tri}(b).~\cite{s617,t1008}
Thus not only the LO but also the TO phonon modes can be observed, 
which is consistent with the previous experimental results~\cite{yu01,b1016} 
and the theories.~\cite{s617,t1008}
Thus when we see the G-band Raman spectra for metallic nanotubes, 
we can expect the following features: 
for armchair nanotubes, we can see both LO and TO in which 
the LO mode appears at a lower frequency region with some spectral width 
while the TO mode appears without spectral broadening because
of the absence of phonon softening. 
For zigzag nanotubes, 
we can see only a broadened LO phonon mode 
but we can not see a TO phonon mode. 
For a general chiral nanotube, 
we generally observe both LO and TO phonon modes
with some broadening as a function of $\theta$. 
Further, we expect some phonon hardening effect for the TO phonon mode. 
The relative intensity between the LO and TO phonon modes is
determined by the chirality dependent
Raman intensity and spectral width. 
The detailed Raman spectral features for all metallic $(n,m)$ nanotubes 
will be presented elsewhere.

The gauge field descriptions 
for the lattice deformation (Eq.~(\ref{eq:A})) and  
for the el-ph interaction of the LO and TO modes
(Eq.~(\ref{eq:optigauge})) 
are useful to show the appearance 
of the curvature-induced mini energy gap  
in metallic zigzag carbon nanotubes and the decoupling
between the TO mode with a finite wavevector and the electrons, as
shown in the following.   
For a zigzag nanotube, 
we have $\delta \gamma^1_0=0$ and 
$\delta \gamma^2_0=\delta \gamma^3_0$ from the
rotational symmetry around the tube axis  (see
Fig.~\ref{fig:graphene}). 
Then, Eq.~(\ref{eq:A}) shows that for $A_x \ne 0$ and $A_y = 0$, 
the cutting line of $k_x=0$ for the metallic zigzag
nanotube is shifted by a finite constant value of $A_x$ because of
the Aharanov-Bohm effect for the lattice distortion-induced gauge
field ${\bf A}$. 
This explains the appearance of the curvature-induced mini energy gap  
in metallic zigzag carbon nanotubes~\cite{kane97} and  
of the phonon broadening for the TO mode as a function of $E_{\rm F}$. 
The TO phonon mode with ${\bf q}\ne 0$ does not change the
area of the hexagonal lattice but instead gives rise to a shear
deformation. Thus, the TO mode (${\bf u}_{\rm TO}({\bf r})$)
satisfies
\begin{align}
 \nabla \cdot {\bf u}_{\rm TO}({\bf r}) = 0, \ \ \ \
 \nabla \times {\bf u}_{\rm TO}({\bf r}) \ne 0.
 \label{eq:TO}
\end{align}
Using Eqs.~(\ref{eq:optigauge}) and (\ref{eq:TO}), 
we see that 
the TO mode does not yield the deformation-induced magnetic field
(${\bf B}({\bf r})=\nabla \times {\bf A}({\bf r})$),
but the divergence of ${\bf A}({\bf r})$ 
instead because
\begin{align}
 \begin{split}
  & B_z({\bf r})= 
  -\frac{g}{v_{\rm F}} \nabla \cdot {\bf u}_{\rm TO}({\bf r})=0,
  \\ 
  & \nabla \cdot {\bf A}({\bf r})=
  \frac{g}{v_{\rm F}}
  (\nabla \times {\bf u}_{\rm TO}({\bf r}))\cdot {\bf e}_z
  \ne 0.
\end{split}
\end{align}
Thus, we can set ${\bf A}({\bf r})=\nabla \varphi({\bf r})$
where $\varphi({\bf r})$ is a scalar function.
Since we can set $ \nabla \cdot {\bf A}({\bf r})=0$ 
in Eq.~(\ref{eq:weyl})
by a redefinition of the phase of the wavefunction
(a local gauge symmetry) 
as $\Psi({\bf r}) \to \exp( -i\varphi({\bf r})/\hbar )
\Psi({\bf r})$,~\cite{sasaki05}
and thus
the ${\bf A}({\bf r})$ in Eq.~(\ref{eq:weyl})
disappears for the TO mode with ${\bf q}\ne 0$.
This explains why the TO mode with ${\bf q}\ne 0$ 
completely decouples from the electrons
and that only the TO mode with ${\bf q}= 0$
couples with electrons.
This conclusion is valid even when the graphene sheet 
has a static surface deformation.
In this sense, the TO phonon mode at the $\Gamma$-point 
is anomalous since the el-ph interaction for the TO mode
can not be eliminated by a phase of the wavefunction. 
It may be difficult to include the local gauge symmetry
in a numerical analysis.
On the other hand, 
the LO phonon mode with ${\bf q}\ne 0$
changes the area of the hexagonal lattice
while it does not give rise to a shear deformation.
Thus, the LO mode (${\bf u}_{\rm LO}({\bf r})$) satisfies
\begin{align}
 \nabla \cdot {\bf u}_{\rm LO}({\bf r}) \ne 0, \ \ \ \
 \nabla \times {\bf u}_{\rm LO}({\bf r}) = 0,
 \label{eq:LO}
\end{align}
Using Eqs.~(\ref{eq:optigauge}) and (\ref{eq:LO}), 
we see that the LO mode gives rise to a 
deformation-induced magnetic field as
\begin{align}
 B_z({\bf r}) \ne 0, \ \ \
 \nabla \cdot {\bf A}({\bf r})= 0.
\end{align}
Since a magnetic field changes the energy band structure
of electrons,
the LO mode (with ${\bf q}\ne 0$) couples strongly to the electrons.

In the case of 2D graphene, 
Eq.~(\ref{eq:coupling}) tells us that 
the $\Gamma$ point TO and LO modes give the same energy shift 
because the integral over $\theta({\bf k})$ gives the same 
$\hbar \omega_\lambda^{(2)}$ in Eq.~(\ref{eq:omega_2})
for both TO and LO modes.
This explains why no $G$-band splitting 
is observed in a single layer of graphene.~\cite{yan07} 
Even when we consider the TO and LO modes 
near the $\Gamma$ point,
we do not expect any splitting between the LO and TO phonon energies
because the TO mode with ${\bf q}\ne 0$
is completely decoupled from the electrons.
Thus, for ${\bf q}\ne 0$, only the LO mode contributes to the $G$-band.


In our numerical results for $\hbar \omega_\lambda$,
we see a crossing between the LO and TO mode 
at some values of $E_{\rm F}$.
If there is a finite transition amplitude between the LO and TO phonon
modes, we need to diagonalize the following matrix
according to second-order perturbation theory 
for a degenerate case,
\begin{align}
 \begin{pmatrix}
  \langle \omega_{\rm LO} | {\cal H}^{(2)} | \omega_{\rm LO} \rangle
  & \langle \omega_{\rm TO} | {\cal H}^{(2)} | \omega_{\rm LO} \rangle \cr
  \langle \omega_{\rm LO} | {\cal H}^{(2)} | \omega_{\rm TO} \rangle
  &\langle \omega_{\rm TO} | {\cal H}^{(2)} | \omega_{\rm TO} \rangle
 \end{pmatrix},
\end{align}
where ${\cal H}^{(2)}$ is the effective Hamiltonian 
in second-order perturbation theory given by
\begin{align}
 {\cal H}^{(2)} = 
 \sum_{\bf k} 
 \frac{{\cal H}_{\rm int} |{\rm eh}({\bf k}) \rangle
 \langle {\rm eh}({\bf k}) | {\cal H}_{\rm int}
 }{\hbar\omega_{\lambda}^{(0)}-(E_{\rm e}({\bf 
 k})-E_{\rm h}({\bf k})) + i\Gamma}.
 \label{eq:d2nd}
\end{align}
Using Eq.~(\ref{eq:coupling}), we see that 
$\cos\theta({\bf k})\sin\theta({\bf k})$
which is an odd function of $k_2$ appears in the numerator
for the diagonal terms in Eq.~(\ref{eq:d2nd}).
Thus, the mode coupling through the electron-hole pair at $k_2$ 
is canceled by that at $-k_2$, i.e., 
due to the time-reversal symmetry of the system.
Thus, even though the LO and TO modes cross each other,
there is no-mixing between the LO and TO phonon modes
for any chirality.
If there is a breaking of time reversal symmetry for $k_2$,
we expect some mixing between the LO and TO phonon modes.

\section{Comparison with experiment and summary}\label{sec:sum}

The first comparison we make is to
experimental results in the literature.
Here we try to assign the chirality of the nanotube 
that is given in Fig.~2 in Ref.~\onlinecite{l1208}.
This figure shows two strong intensity peaks 
with different phonon frequencies.
The higher frequency peak does not depend on the gate voltage 
and the lower frequency peak shows a frequency shift 
with a strong broadening near the Dirac point.
The existence of a flat intensity peak 
as a function of Fermi energy for a metallic SWNT 
indicates that the electron-phonon coupling is very weak
for the phonon mode.
Thus, the nanotube can be thought of as an armchair SWNT
(or close to an armchair SWNT in the chiral angle)
since the el-ph coupling between the TO mode 
and electron-hole pairs is negligible in this case.
Further investigation comparing theory with experiment
is strongly necessary for tubes of different chiralities.

\begin{figure}[htbp]
 \begin{center}
  \includegraphics[scale=0.45]{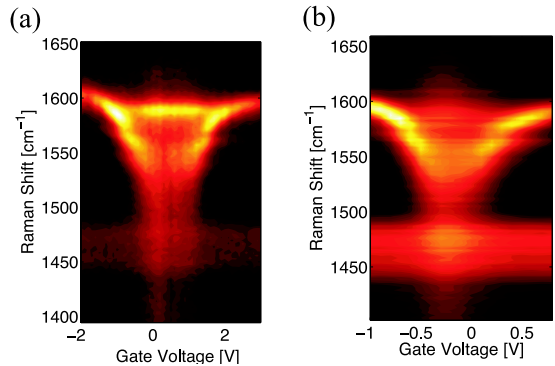}
 \end{center}
 \caption{(color online)
 Experimental results of the Raman G-band intensity 
 as a function of applied gate voltage for two metallic SWNTs (a) and (b).
 A strong (weak) intensity peak is denoted by the yellow (black) color.
 }
 \label{fig:jingkong-data}
\end{figure}

In Fig.~\ref{fig:jingkong-data}(a) and (b), we
show experimental results of the G-band intensity
as a function of applied gate voltage for two
different isolated SWNTs.  
For the given excitation energy of 1.91eV and 
the observed radial breathing mode (RBM) frequencies 196 and 193
cm$^{-1}$ for the sample of
Fig.~\ref{fig:jingkong-data}(a) and (b),
respectively,
we can assign $(n,m)$ by using the conventional assignment
technique that we used for single nanotube Raman
spectroscopy.\cite{l818}
As a result, we assign the $(n,m)$
values as (12,6) for Fig.~\ref{fig:jingkong-data}
(a) and (15,3), (16,1), or (11,8) for
Fig.~\ref{fig:jingkong-data}(b).

Although (12,6) is not so close to the chiral
angle for the armchair nanotube, we expect that
Figure~\ref{fig:jingkong-data}(a) exhibits a
similar behavior to that of Fig.~2 in
Ref.~\onlinecite{l1208}. Thus the assignment of
(12,6) from the RBM spectra is consistent with the
present phonon spectra.  
In Fig.~\ref{fig:jingkong-data}(b), the intensity
near the Dirac point of the TO (LO) mode is weaker
(stronger) than shown in (a).  
This indicates that 
the tube is not an armchair SWNT 
but close to a metallic zigzag SWNT 
(or a metallic chiral SWNT). 
Thus we can assign this SWNT either to (15,3)
or to (16,1) for the SWNT for
Fig.~\ref{fig:jingkong-data}(b)
and we can exclude $(11,8)$.  
A further comparison with more experimental data 
will be required, 
which will be carried out in the future.

A chirality dependent Raman intensity 
was observed for metallic SWNTs
as shown at Fig.~1 in Ref.~\onlinecite{wu07}.
Although the $E_{\rm F}$ positions for 
the observed metallic SWNTs are unclear,
the results are consistent with our calculations
in the following way.
For example, 
the TO mode in a $(15,15)$ SWNT 
gives a sharp lineshape and
the LO mode shows a broad feature for a $(24,0)$ SWNT.
It is noted that the Raman intensity
is proportional to the el-ph coupling for an 
optically excited electron via Eq.~(\ref{eq:coupling2}).
Since this el-ph coupling depends on the chiral angle due to 
the trigonal wrapping effect,
one of the two optical modes may be invisible
in the Raman intensity.~\cite{wu07,reich04}

In summary, 
we have calculated the $E_{\rm F}$ dependence of $\hbar \omega_\lambda$
($\lambda={\rm LO}$ and TO) for metallic carbon nanotubes.
The results show a strong dependence of the phonon frequency shift 
on the chirality of single-walled carbon nanotubes 
because of the curvature-induced 
shift of the wavevector around the tube axis.
This is explained by the general property of 
second-order perturbation theory 
and the characteristic electron-phonon coupling for 
the $\Gamma$ point LO and TO phonon modes 
(Eqs.~(\ref{eq:coupling}) and (\ref{eq:H_int})).
For the LO and TO phonon modes near the $\Gamma$ point,
we showed the absence of an electron-phonon coupling 
for the TO mode for ${\bf q}\ne 0$
due to a local gauge symmetry, 
and that the LO mode works as 
a deformation-induced magnetic field.
The phonon mixing between LO and TO phonon modes
is absent in second-order perturbation theory
due to time-reversal symmetry.

\section*{Acknowledgment}

R. S. acknowledges a Grant-in-Aid (No. 16076201) from MEXT.
MIT authors acknowledge support under NSF Grant
DMR 07-04197.


\end{document}